\documentclass[thesis]{MSUstyle}

\usepackage[final]{graphicx} 
\usepackage{subcaption}    
\usepackage{color} 
\usepackage{varioref}  
\usepackage[final]{listings} 
\usepackage[pdftex,hidelinks]{hyperref}
\usepackage[numbers,sort&compress]{natbib} 
\usepackage{longtable} 
\usepackage{natbib}
\usepackage{tikz} 
\usetikzlibrary{calc}
\usetikzlibrary{patterns}
\usetikzlibrary{positioning}
\usetikzlibrary{decorations.markings}
\usetikzlibrary{decorations.pathmorphing}
\usetikzlibrary{arrows}
\usepackage{accessibility}

\usepackage{amsmath}
\usepackage{lipsum} 

\name{Zachary Douglas Wadhams} 
\DocTitle{On improving the adoption, usability, and retention of Static Application Security Testing (SAST) tools} 
\OWNwebpage{www.linkedin.com/in/zachary-wadhams-452305192} 
\degreetitle{Cybersecurity} 
\department{Computer Science} 
\committeechair{Dr. Clemente Izurieta} 
\departmentchair{Dr. John Paxton}
\graduatedean{Dr. Craig Ogilvie}
\submitdate{December 2024} 
\copyrightyear{2025} 

\keys{Cybersecurity, Computer Science, Masters, Thesis, Static Analysis, SAST, Static Application Security Testing, Vulnerability Detection, Montana, Montana State University} 

\hypersetup{%
	baseurl={\OWNwebpage}, 
	citecolor=black, 
	colorlinks=true, 
	draft=false, 
	filecolor=black, 
	linkcolor=black, 
	pdfauthor={\name}, 
	pdfcreator={\name \ by \ PDFLaTex}, 
	pdfdisplaydoctitle=true, 
	pdffitwindow=true, 
	pdfkeywords={\keys}, 
	pdfsubject={\degreetype \ for \name}, 
	pdftitle={\DocTitle}, 
	plainpages=false, 
	urlcolor=black 
}

\begin{document}

\begin{preliminary} 

  \begin{acknowledgements}

I would like acknowledge and thank my advisor and committee chair, Dr. Clemente Izurieta for his unwavering support throughout my program. I would also thank committee members Dr. Ann Marie Reinhold and Dr. Matthew Revelle and the entirety of the MSU Software Engineering and Cybsersecurity Lab (SECL) for always being on my side. And finally, my gratitude to Clara for pushing me further than I had ever thought possible.

This research was supported by TechLink (TechLink PIA FA8650-23-3-9553). Any opinions contained herein are those of the author and do not necessarily reflect those of TechLink. 
  \end{acknowledgements}

  \contents

  \begin{abstract}
As the Internet connects our world ever closer and propels human progress toward new frontiers, it also exposes us to new and unforeseen dangers. Now that the majority of humanity is connected to the Internet, bad actors can potentially reach millions of people with the press of a button. With software being the primary medium on which the Internet is used, and with many Internet security breaches resulting from code vulnerabilities, a level of security is necessitated. The responsibility of securing these applications falls on the software developers. Fortunately, a variety of tools and techniques exist to assist developers in identifying and resolving software vulnerabilities. Static Application Security Testing (SAST), one of these tools, employs automated analysis techniques to meticulously examine an application's source code. This examination occurs early in the development process, even before the code is functional. SAST tools pinpoint potential security weaknesses within the code's structure, highlighting areas where malicious actors might exploit vulnerabilities. By identifying these risks early on, SAST tools allow developers to proactively address security concerns and build more robust applications. Despite these benefits, SAST tools are far from perfect. Our research focuses on challenges developers encounter when using these tools, with the overarching goal being to improve its usability. We first present a literature review that examines 89 works of research relating to the implementation and continued usage of SAST. Through this review, we uncovered various problems developers had with SAST. Some of these, such as false positives, which are security warnings that identify a potential vulnerability that doesn't actually exist in the code, were mentioned in a majority of the 89 papers we reviewed. The second manuscript details a process for automating the execution SAST tool output with a focus on presenting the data in a format that is meaningful and actionable to developers. This includes a real world use case example that provided feedback on an implementation of our process. Developers indicated satisfaction with many aspects of the process and conveyed that it made them more willing to use the SAST tool.

  \end{abstract}
\end{preliminary}
\chapter{Introduction}\label{CH:introduction}

Humanity has crossed a threshold, and there is no going back. We have allowed software to infiltrate nearly all aspects of our professional and personal lives, to great benefit of productivity, accelerating innovation, and connecting people across the globe. However, in doing so we have exposed all which that software touches. Be it our personal intimate moments or our professional careers, malicious actors have the potential to access and expose it all. The only thing standing between cybercriminals and users is the security and robustness of software. 

In 2023, there were a staggering 3,205 reported data compromises, affecting over 350 million individuals. 73\verb|%| of these compromises resulted from cyberattacks. Compared to the all-time high reported in 2021, the number of compromises has surged by  72\verb|%| \cite{ITRC2024}. The cost of these attacks in 2023 has been estimates to cost over 8 trillion USD globally \cite{Morgan2023}. These data points paint a clear picture: cybercrime is a lucrative and persistent threat that will only grow over time.

The responsibility to defend against these threats ultimately falls on the organizations that develop software. Fortunately, software developers have several tools and techniques to secure applications. Penetration testing or pen testing involves a cybersecurity expert probing around in your system to attempt to exploit vulnerabilities and gain unwanted access \cite{CloudflarePen}. Essentially, they play the role of a bad actor to simulate potential ways of entry or disruption. Dynamic Analysis or Dynamic Application Security Testing (DAST) simulates an application's runtime behavior against attack scenarios \cite{IBMDAST}. Both penetration testing and DAST require a functional software application to effectively uncover security vulnerabilities. This presents a challenge for applications in their infancy that may not be fully fleshed out or executable. In the absence of a running application, however, source code remains a valuable asset.

The work of this thesis focuses on another method, Static Application Security Testing (SAST), also known as static analysis. SAST is a method for analyzing source code to identify potential security vulnerabilities without requiring a fully functional application \cite{SonarSourceSAST}. By inspecting the codebase early in development, SAST assists developers in proactively addressing security risks, preventing them from becoming more complex and harder to mitigate later in the development process. 

Prior research in the field has shown the benefits of SAST. Chess et al. \cite{Chess2004} demonstrates how static analyzers save developer time in reviewing code for bugs and vulnerabilities when compared to a manual review. Additionally, these tools also cover a broader scope that a human ever could. Yang et al. \cite{Yang2019} conducted a study showing that when taking the advice of a SAST tool, Priv, versus the advice of a security expert, the suggested fix was identical 75.3\verb|%| of the time. The SAST tool was also able to provide a complete and correct fix for 75.6\verb|%| of the warnings. 

Notwithstanding these benefits, some development teams either have chosen not to use SAST or have used it at some point and then either partially or wholly abandoned it. There are some studies that have taken different approaches in trying to identify the reason behind this conundrum. In 2013, Johnson et al. \cite{Johnson2013} interviewed 20 software developers whose years of development experience ranged from 3 to 25 years. The average experience of these developers was 10 years. Johnson's study suggested that false positives were a major issue when it came to a development team considered using SAST. A false positive in SAST occurs when the tool erroneously identifies code as containing a vulnerability. A developer will investigate this issue, only to find that it does not apply to their project. This wastes developer work time and can cause frustration at the tool. Johnson additionally identified lack of comprehensive integration into workflow and insufficient explanation of defects as problems developers cited. Nachtigall et al.'s \cite{Nachtigall2022} study presents a novel approach to SAST usability evaluation. Employing a comprehensive set of 36 criteria, the research assessed 46 SAST tools across various dimensions, including warning messages, fix support, false positives, and workflow integration to name a few. This substantial undertaking involved the setup, execution, and analysis of each tool. Nachtigall's methodology allowed them to gain a personal understanding of the types of problems developers may encounter when using these tools. Specifically, they showed that over half the tools had warning messages with insufficient detail and over three-quarters had poor fix suggestions. Also, many tools completely ignored incorporating user knowledge. For example, a prime technique to cut down on the rate of false positives is to allow users to flag an issue as such. The next time the tool is run, it will see that flag and then not present that reported false positive to the developer. They also identified issues with cohesive workflow integration and various other problems. While both of these studies are eye opening and greatly beneficial to the field of SAST, they stop at identifying the issues and don't provide any detailed solutions or improvements.

\begin{centering}
\begin{figure*}[!h]
\centerline{\includegraphics[scale=.35]{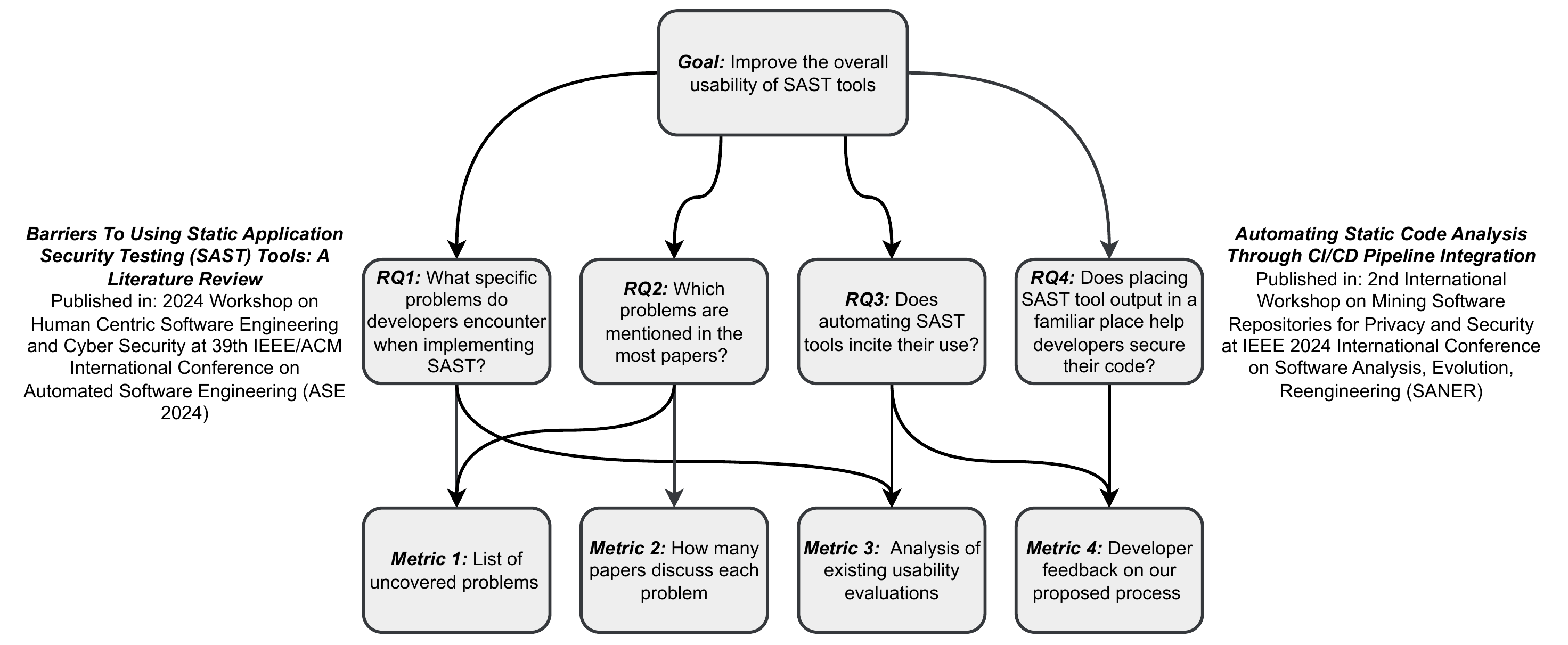}}
\caption{Hierarchical Goal-Question-Metric(GQM) structure based on Basili's Goal-Question-Metric methodology \cite{Basili1994}. Each of the research questions is addressed by a specific manuscript. Research questions 1 and 2 correspond to the manuscript on the left while questions 3 and 4 correspond to the manuscript on the right.}
\label{gqm}
\end{figure*}
\end{centering}

By comprehensively examining existing research and conducting practical experimentation, this thesis aims to bridge the gap between the identification of SAST usability challenges and the implementation of effective solutions. This will be achieved through two primary steps. First, we will confirm previously identified SAST usability challenges, identify new issues, and develop potential solutions for each. Second, we will implement one of these proposed solutions in a real-world setting to evaluate its efficiency and effectiveness. To accomplish this, we leveraged Basili's Goal-Question-Metric (GQM) \cite{Basili1994} approach, outlined in Figure \ref{gqm}. This framework guided the development of four Research Questions, two for each manuscript. Each Research Question is linked to one or more metrics, as shown in Figure \ref{gqm}.

The first manuscript presented, 'Barriers to Using Static Application Security Testing (SAST) Tools: A Literature Review', is an in depth analysis of 89 papers in the SAST space published in the last 5 years. The specific target of this literature review was usability challenges and problems related to its implementation and continued usage. The results from this manuscript directly informed Research Question 1 (What specific problems do developers encounter when implementing SAST?) and Research Question 2 (Which problems are mentioned in the most papers?), as seen in Figure \ref{gqm}. Research Question 1 is answered by metrics 1 and 3, while Research Question 2 is answered by metrics 1 and 2. 

The next and final manuscript, 'Automating Static Code Analysis Through CI/CD Pipeline Integration', presents a novel, generalized, and automated process for integrating SAST tools into a developer's familiar issue tracking software. Crucially, this paper involves a real-world use case in an active development environment. The feedback provided by developers who used this process informed us of its effectiveness and whether it encouraged developers to consistently use SAST. This manuscript asks Research Questions 1 and 2. Research question 1 is answered by metrics 3 and 4, and research question 4 is addressed by metric 4. 

The following sections will present our findings from a rigorous literature review and a real-world implementation of a SAST integration solution.
\chapter{Research Goals}\label{CH:goal}

\section{Goal Question Metric}
To ensure a well-structured research plan, we employed Basili's Goal-Question-Metric approach for defining our overarching research goal; improve the overall usability of SAST tools \cite{Basili1994}. To address this goal, we created four research questions, each answered by one or more metrics. The structure of this plan is detailed in Figure \ref{gqm} and in the following text.

{\large \textbf{Goal:} Improve the overall usability of SAST tools.}

\begin{itemize}
    \item \textbf{Research Question 1:} What specific problems do developers encounter when implementing SAST?
    \begin{itemize}
        \item \textbf{Metric 1:} List of uncovered problems
        \item \textbf{Metric 3:} Analysis of existing usability evaluations
    \end{itemize}
    \item \textbf{Research Question 2:} Which problems are mentioned in the most papers?
    \begin{itemize}
        \item \textbf{Metric 1:} List of uncovered problems
        \item \textbf{Metric 2:} How many papers discuss each problem
    \end{itemize}
    \item \textbf{Research Question 3:} Does automating SAST tools incite their use?
    \begin{itemize}
        \item \textbf{Metric 3:} Analysis of existing usability evaluations
        \item \textbf{Metric 4:} Developer feedback on our proposed process
    \end{itemize}
    \item \textbf{Research Question 4:} Does placing SAST tool output in a familiar place help developers secure their code?
    \begin{itemize}
        \item \textbf{Metric 4:} Developer feedback on our proposed process
    \end{itemize}
\end{itemize}


\chapter{Barriers to Using Static Application Security Testing (SAST) Tools: A Literature Review}\label{CH:manuscript1}

\section{Contribution of Authors and Co-Authors}

Manuscript in following chapter

Author: Zachary Wadhams
\begin{singlespace}
Contributions: Developed study concept and design, data collection and analysis,\\\indent interpretation of results, and wrote the manuscript.
\end{singlespace}

Co-Author: Clemente Izurieta
\begin{singlespace}
Contributions: Obtained funding, provided feedback and editing.
\end{singlespace}

Co-Author: Ann Marie Reinhold
\begin{singlespace}
Contributions: Obtained funding, provided feedback and editing.
\end{singlespace}


\newpage

\section{Manuscript Information}

Zachary Wadhams, Ann Marie Reinhold, Clemente Izurieta

\begin{singlespace}
2024 Workshop on Human Centric Software Engineering and Cyber Security, HCSE\verb|&|CS-2024 (ASE 2024)
\end{singlespace}

\begin{singlespace}
  Status of Manuscript: \\\indent 
  \underline{\phantom{~~X~~}} Prepared for submission to a peer-reviewed journal\\\indent
  \underline{\phantom{~~X~~}} Officially submitted to a peer-reviewed journal\\\indent
  \underline{\phantom{~~X~~}} Accepted by a peer-reviewed journal\\\indent
  \underline{~~X~~} Published in a peer-reviewed journal
\end{singlespace}

\begin{singlespace}
\end{singlespace}

\newpage

\section{Abstract}
Developers face a challenging problem with no clear solution. Modern software breaches can wreak havoc on businesses and individuals alike. With code vulnerabilities being a leading cause, securing applications must be a priority for developers. Static Application Security Testing (SAST) has the potential to harden applications by assisting in the identification and resolution of security vulnerabilities. Despite this, many development teams have not adopted SAST tools into their environment. In this paper, we survey the recent literature to uncover why some developers are apprehensive towards SAST and identify what specific problems they encounter when using it. We found a variety of usability problems developers face when using SAST. Some are inherent of the tool and ultimately require some level of developer investment while others are tool shortcomings that SAST tool creators must address. Ultimately, we argue that in order to drive widespread adoption and consistent SAST usage, developers will need to embrace that some investment is required. Simultaneously, developers will be more likely to integrate SAST tools into their workflows if the creators of SAST tools simplify many aspects related to tool usage. Surmounting the primary obstacles preventing the adoption of SAST requires full consideration of both the technical and human factors.
\newpage

\section{Introduction}
Recent years have witnessed a surge in critical software security issues, impacting millions of people and causing billions of dollars in damages \cite{Alkhadra2021}.  In July of 2024, a faulty CrowdStrike update unintentionally crippled Windows systems globally, highlighting the far-reaching consequences of software defects \cite{CrowdStrike2024}. The 2020 SolarWinds attack stands as a well known example, where a code vulnerability allowed attackers to inject malicious code into software updates. This breach exposed sensitive data, disrupted critical infrastructure, and is estimated to have cost \verb|$|100 billion to recover, impacting countless individuals and businesses \cite{Thomas2018}. These incidents underscore the critical need for robust secure coding practices throughout the software development lifecycle.

Fortunately, there are myriad of tools and techniques developers can take advantage of to secure their codebases. These include Dependency Scanning, Penetration Testing, Dynamic Code Analysis and Static Application Security Testing (SAST), among others \cite{ODonoghue2024, Bertoglio2017, Lindlan2000}. Static Application Security Testing, in particular, is unique as it can be implemented at any time in the development lifecycle to identify and help to resolve vulnerabilities. It can achieve this because of how it works--by scanning source code without it needing to run or compile \cite{Yang2019}. Such benefits have been recognized by Ayewah et al., who showed that many overlooked vulnerabilities were resolved after developers were alerted by warnings from SAST tools \cite{Ayewah2010}. 

While implementing SAST is relatively easy and has shown benefits, some development teams still choose not to use it, opting for limited code analysis, focusing on security-critical components only, or relying mostly on manual code reviews \cite{Johnson2013, Lopez2023, Xie2011}. This study aims to identify the root causes of developer apprehension regarding full SAST adoption. We focus on specific issues that discourage developers from initiating or abandoning SAST use. By understanding these challenges, we hope to shed light on areas for improvement in SAST usability and encourage further research directed at enhancing the developer experience.

\section{Related Work}
Previous surveys and studies have explored usability challenges associated with SAST tools. For instance, Johnson et al. \cite{Johnson2013} interviewed 20 software developers to understand their perspectives on usability issues. Charoenwet et al. \cite{Charoenwet2024} examined SAST tools for their effectiveness in security code reviews. In contrast, our study aims to gain a broad understanding of the current state of SAST usability through a comprehensive literature review.

\section{Methodology}
\label{sec:Methodology}
An overview of our study methodology can be found in Figure 1 and we will reference it throughout this section. The initial phase of our study involved choosing the databases from which to gather papers. We chose the ACM Digital Library and IEEE Xplore due to their extensive collections of peer-reviewed research articles and conference proceedings in the field of software security and development. 

\begin{figure*}[t]
\centerline{\includegraphics[scale=0.2]{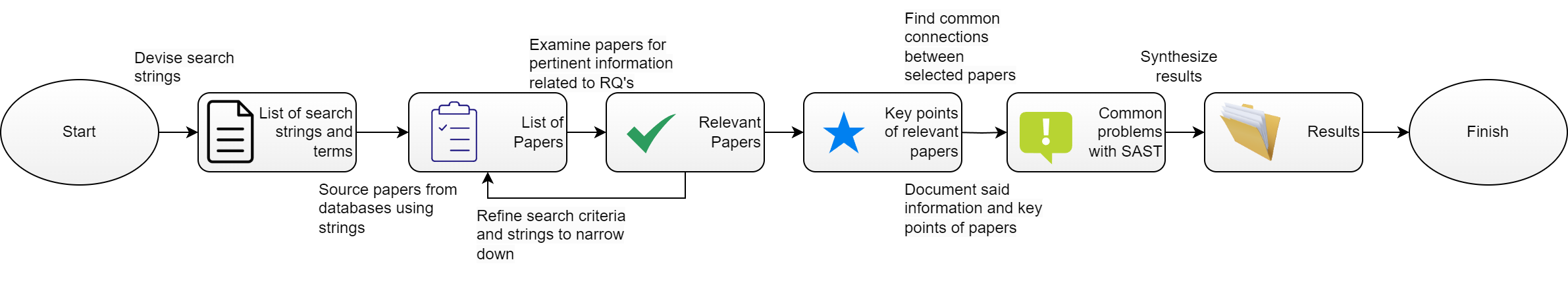}}
\caption{Study Method Diagram. Search strings were devised and refined. Following refinement of strings, all 240 papers from the ACM Digital Library and all 108 papers from IEEE Xplore were examined for relevancy and either discarded or kept in accordance with our inclusion criteria. All relevant papers were read from cover-to-cover, thoroughly examined, and the issues developers encountered were cataloged. We then examined these issues systematically.}
\alt{Study Method Diagram. Search strings were devised and refined. Following refinement of strings, all 240 papers from the ACM Digital Library and all 108 papers from IEEE Xplore were examined for relevancy and either discarded or kept in accordance with our inclusion criteria. All relevant papers were read from cover-to-cover, thoroughly examined, and the issues developers encountered were cataloged. We then examined these issues systematically.}
\label{fig:Figure1}
\end{figure*}

\begin{figure}[t]
\centerline{\includegraphics[scale=0.55]{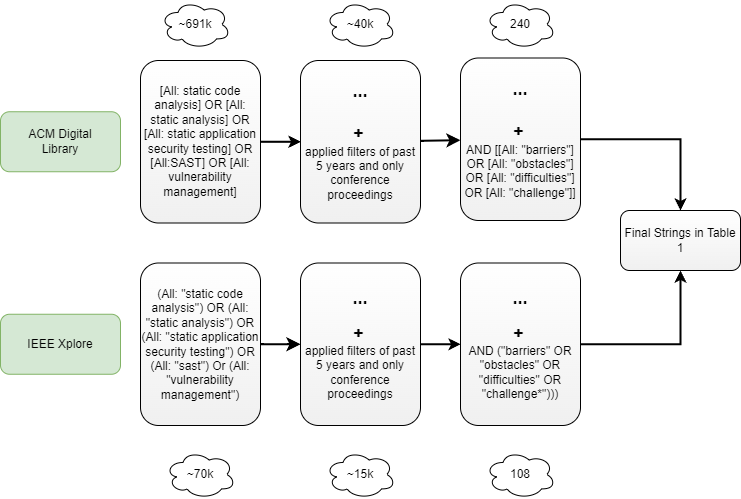}}
\caption{Search String Refinement. This figure shows our search strings, refinement process, and the number of papers returned at each step of the process. We denote each refined string with ``… +" to indicate that all the string everything after the + was concatenated to the string(s) indicated in the boxes to the left.}
\alt{Search String Refinement. This figure shows our search strings, refinement process, and the number of papers returned at each step of the process. We denote each refined string with ``… +" to indicate that all the string everything after the + was concatenated to the string(s) indicated in the boxes to the left.}
\label{fig:Figure2}
\end{figure}

Next, we devised initial search strings as shown in the first step in Figure \ref{fig:Figure1}. These strings provided an estimate of the number of papers relating to our topic. Figure \ref{fig:Figure2} displays these strings and the results of searching on them. The initial strings returned thousands of papers from each database. Consequently, we refined the strings, as detailed in Figure \ref{fig:Figure2}, until we reached a manageable number of papers. This is represented by the looping section from the clipboard to the check mark in Figure \ref{fig:Figure1}.

Our final search strings returned 240 papers from the ACM Digital Library and 108 papers from IEEE Xplore. Upon further examination of the results from the ACM Digital Library, we noted that the Brazilian Symposium of Systematic and Automated Software Testing \footnote{https://dl.acm.org/conference/sast} introduced 118 non-relevant papers into the 240 that were sourced. This inflation occurred because the acronym ``SAST" was used in all publications from this conference, despite these papers not covering SAST. Removal of these conference papers reduced the number of ACM sources to 122 papers.

We then manually reviewed each of these 230 papers (108 from IEEE Xplore and 122 from ACM Digital Library sans the Brazilian Symposium of Systematic and Automated Software Testing papers) for their relevance. For the first pass, we specifically looked for papers that mentioned an implementation of SAST. We then realized that many papers mention static analysis or SAST in a general sense without any relation to the actual usage of these techniques. These papers were discarded. After the initial review, we identified a total of 89 relevant papers.

After identifying these papers, we analyzed them to identify key points related to SAST and developer usage, particularly the challenges encountered, shown at the star in Figure \ref{fig:Figure1}. We documented these problems and noted the number of papers in which they occurred. 

\begin{figure}[h]
        \centering
        \includegraphics[scale=0.55]{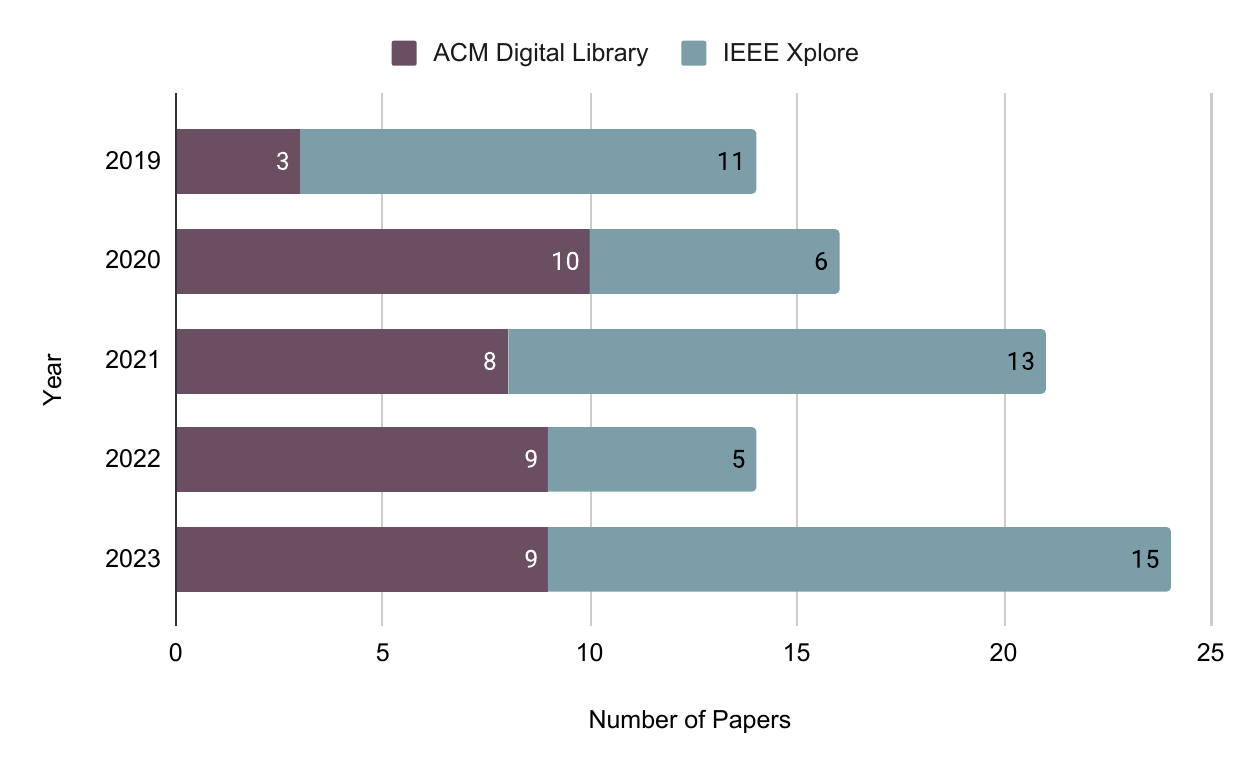} 
        \caption{Temporal Publication Trends in SAST Conference Proceedings. The ACM Digital Library is depicted in purple and IEEE Xplore in blue. An increasing trend over time can be seen, with the highest number of papers published in 2023.}
        \alt{Temporal Publication Trends in SAST Conference Proceedings. The ACM Digital Library is depicted in purple and IEEE Xplore in blue. An increasing trend over time can be seen, with the highest number of papers published in 2023.}
        \label{fig:Figure4}
\end{figure}

\section{Results and Discussion}
\label{sec:Results and Discussion}
In general, the number of publications increased annually from 2019 - 2023 (Figure 3). Except for a slight drop in 2022, there is a year-by-year increase in research on SAST implementation. This trend suggests that research interest in the use of SAST is increasing.

Many papers shared the following pattern. First, a tool would be selected and set up according to an organization’s environment. These tools then require some level of attention to maintain which increases the effort needed from developers. As a result, SAST tools become cumbersome to use and thereby less attractive to those developers. Some papers suggest that these developers gradually reduced their use of the tools until they no longer served their original purpose and became defunct. Within organizations, the decline in usage can be attributed to several usability challenges (Figure \ref{fig:Figure5}), discussed in the following subsections.

\subsection{False Positives}
\begin{figure}[t]
\centerline{\includegraphics[scale=0.55]{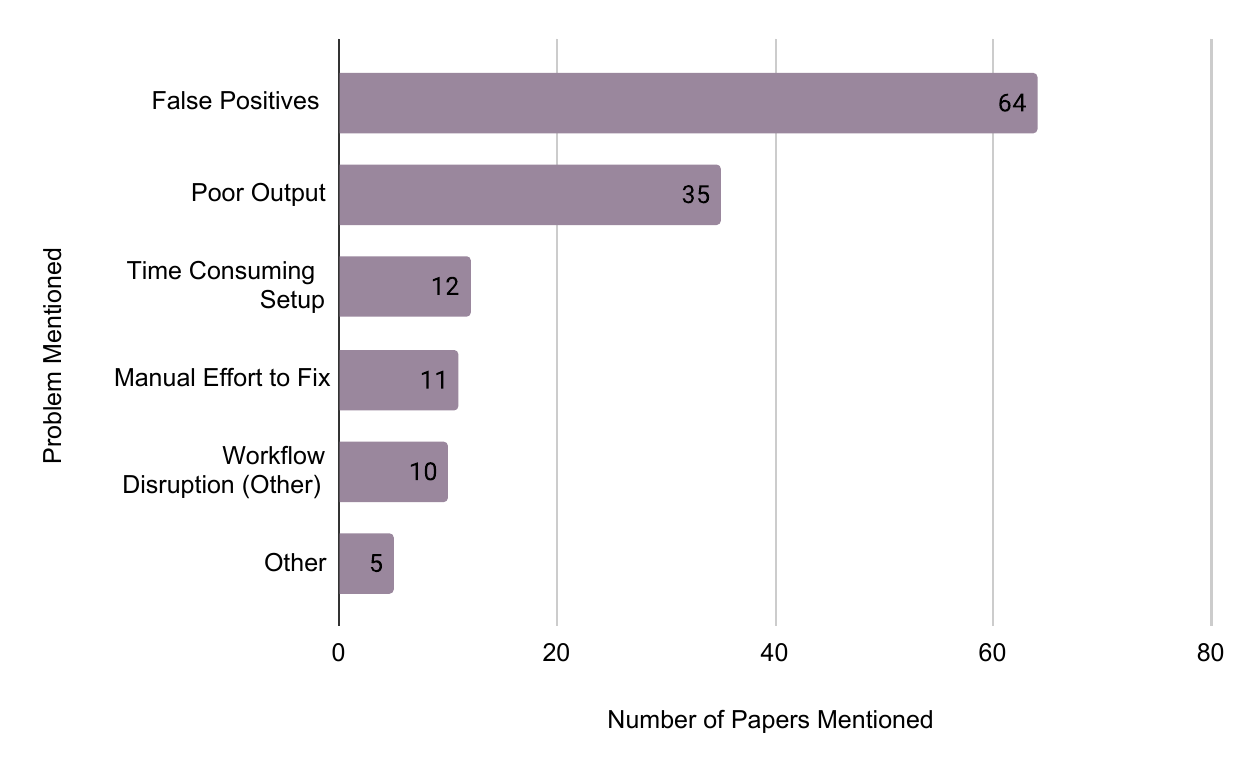}}
\caption{Number of papers in which SAST related problems are mentioned. This figure shows the number of papers each issue was found in. False positives have nearly twice as many occurrences as the next highest complaint: poor output. Time consuming setup and manual effort to fix have a similar amount of instances. Workflow disruption (other) is a grouping of other workflow disruption issues. The final category, other, contains less frequently mentioned issues that were encountered.}
\alt{Number of papers in which SAST related problems are mentioned. This figure shows the number of papers each issue was found in. False positives have nearly twice as many occurrences as the next highest complaint: poor output. Time consuming setup and manual effort to fix have a similar amount of instances. Workflow disruption (other) is a grouping of other workflow disruption issues. The final category, other, contains less frequently mentioned issues that were encountered.}
\label{fig:Figure5}
\end{figure}

In the context of SAST tools, a false positive refers to a situation where the tool incorrectly flags a piece of code as containing a security vulnerability or issue when, in fact, there is no real vulnerability present. Over two-thirds of papers cited false positives as a recurring pain point for developers (Figure \ref{fig:Figure5}). 

These false positives can negatively impact developers in a variety of ways. A primary way this can manifest is by wasting the developer's time \cite{Pan2019}. When a developer has to spend hours tracking down a reported issue only to find that it wasn’t an issue, it can have a cascading result. This wasted time can lead to a decreased trust in the tool, general frustration, and overall reduced productivity \cite{Nachtigall2022}. 

The pervasiveness of false positives necessitates methods to mitigate them. Guo et al. states that it is imperative to implement strategies that minimize their occurrence and impact on the development process \cite{Guo2023}. Such strategies are an open area of research. 

The most common of these strategies is manual review, which requires a developer to investigate each reported issue to determine its validity \cite{Aloraini2019}. If an issue is found to be a false positive, its identifier is added to an ignore list, and the SAST tool will no longer flag it.

\subsection{Poor Output}
Many SAST tools convey their results in a consistent format, either XML or JSON \cite{Johnson2013}. While this is useful for compatibility, developers are seldom interested in the sparse output and minimally formatted text that XML or JSON provides \cite{Nachtigall2022}. Depending on the size of the analysis files can contain thousands of lines and can be cumbersome and overwhelming \cite{Schnappinger2019}. 

A potential strategy to enhance the readability of outputs would involve constructing a parser that formats the issues. While a small number of tools have a built-in parser, the vast majority do not \cite{Wadhams2024}. Adding a parser to facilitate consistent and familiar presentation has been shown to capture developer attention and potentially promote more consistent tool usage \cite{Wadhams2024}. This additional effort could be regarded as addressing the next issue developers may encounter: time-consuming setup.

\subsection{Time-Consuming Setup}
While some SAST tools may be relatively simple to run-—e.g., just point them at a file structure—-others require an intensive setup that demands hours of investment \cite{Morales2020}. This time investment may not be perceived as worthwhile, resulting in a lower adoption rate. A developer might spend hours attempting to configure a tool to run correctly within their environment, only to give up or deem it unworthy of the effort. Thus, time-consuming setup discourages uptake. We encourage the creators of these SAST to address this issue and--to the extent possible--provide clear, comprehensive instructions.

\subsection{Manual Effort}
When a developer uses a SAST tool, automatic code-fix suggestions may be appealing and encourage continued usage of the tool \cite{Odermatt2022}. In contrast, when a tool alerts a developer without providing suggested fixes or a button for automatic correction, they can become frustrated by the manual effort required. This overall lack of guidance was a recurring theme relating to the manual effort required when using SAST tools. 

A potential solution for the lack of guidance is offered by Linters, a static analysis tool that typically integrates into a developer's IDEs and provides real-time feedback \cite{Luo2021}. Additionally, and as a bonus, many Linters have the auto-fix functionality  that developers desire \cite{Habchi2018}. Thus, Linters have the potential to reduce the manual effort required.

\subsection{Workflow Disruption}
Static Application Security Testing can add time to a developer's workflow \cite{Nachtigall2022}, potentially disrupting it. Disrupting a developer's workflow can have significant drawbacks, including decreased productivity, increased frustration, and potential resistance to adopting security practices \cite{Nachtigall2022}. When developers are interrupted by cumbersome tools, it can hinder their workflow and impede their progress on projects \cite{Nachtigall2022}. Additionally, disruptions may lead to context switching, where developers must shift their focus away from coding to address security issues, resulting in loss of momentum and increased cognitive load \cite{Graziotin2017}. This can ultimately impact the quality and timeliness of software delivery. Therefore, minimizing disruptions to a developer's workflow is essential for maintaining productivity.

In order to achieve this, it is important for tools to operate swiftly, ideally to provide timely feedback on a developer's work without prolonged delays. Maintaining rapid software deployment speed is paramount in the implementation of SAST \cite{Rajapakse2021}. Any delay in these deployments, no matter how brief, may lead to skepticism regarding a tool's effectiveness, should the slowness of a tool be identified as the cause.

\subsection{Other Problems}
Various other one-off problems were mentioned in the corpus of papers we evaluated. The first of these is the lack of customizability of many tools \cite{Granata2022}. This lack of customizability can manifest as a barrier in a few ways, with one being difficulty integrating smoothly with a development environment. With each environment being different and custom-tailored to the organization, these tools should strive to easily integrate with as wide of an audience as possible. The other way a lack of customizability can show up is in the actual running of the tool. Some may support a limited number of languages, scale poorly, or not allow users to customize their rules. 

Another mentioned issue is unvalidated metrics \cite{Wyrich2021}. Ideally, the results of a tool should be transparent and trustworthy, yet in many SAST tools, there is no level of proof to support them. Through the phenomena of “new version new answer”, even something as simple as an updated tool version can give wildly different and unexpected results \cite{Reinhold2023}. Because changes based on SAST results can have a significant impact, ensuring their accuracy would build trust in the tool and likely increase usage. 

\subsection{What's Next? Is It All Worth It?} 
Implementing SAST into a development environment helps secure code bases, and fosters more security-conscious developers \cite{Ayewah2010}. Moreover, SAST tools can be efficient in finding and helping resolve security vulnerabilities \cite{Jovanovic2006, Livshits2005}. SAST tools can be viewed as investments, with the payout being a more robust and secure code base in a shorter time period than a manual review could provide. For instance, consider false positives. In the previously mentioned approach, detected false positives are reported to prevent them from appearing in future scans. With repeated implementation over time, a balance may be attained, with improved handling of false positives (e.g., databases containing known false positives). User investment in this effort ameliorates this problem. 

Similarly, writing a parser to convert poor tool output to a better format can enhance a tool’s usability \cite{Wadhams2024}. While there was certainly a significant engineering effort involved, the firsthand developer feedback we received indicated a level of satisfaction with the tool that they did not have before the parser was implemented.

Some barriers to SAST implementation can be overcome through SAST creators developing auto-fix capabilities. Marcilio et al. \cite{Marcilio2019} and Odermatt et al. \cite{Odermatt2022} mentioned that auto-fix capabilities improved the overall usability and integration of tools into workflows and pipelines.

These observations suggest that both developers and SAST tool creators play independent but equally important roles in driving SAST adoption. This paper highlights the substantial advantages for developers who implement SAST proactively. In parallel, SAST tool creators must be responsive to user feedback to ensure continuous improvement in usability.

The successful implementation of SAST tools into development workflows incorporates technological advancements and is also heavily reliant on human factors. We know that the development community must invest significant time and effort to understand the tool's capabilities, interpret results, and address identified vulnerabilities. To further drive SAST adoption, a deeper understanding of the human element is needed. Further research should focus on identifying the institutional, social, and cognitive barriers that hinder SAST usage, such as skepticism about tool accuracy, or resistance to change. We assert that the future of the SAST field hinges on our ability to enumerate and address the human factors impacting developers.

\section{Threats to Validity}
\label{sec:Threats to Validity}
This study offers valuable insights that contribute to the field of SAST and, by extension, secure software development. This section discusses potential threats to the validity of our study. By identifying these threats, our goal is to enhance the study's reliability and reduce the risk of bias.

Firstly, we followed a structured process for identifying relevant papers (Figure 1). Developing this process was the first action we took, therefore ensuring consistency through out each step of the literature review. Although we opted for a structured search strategy tailored to our research questions rather than a predefined protocol like Kitchenham's \cite{Kitchenham2007}, this systematic approach ensured a rigorous selection of relevant literature. While our stringent search strings might have excluded some publications, we believe the resulting sample of 89 papers provides a sufficient window into the current state of the SAST field.

Finally, we opted to manually review all 89 selected papers instead of using automated approaches like Natural Language Processing (NLP). Some may claim that the lack of an automated aspect can limit the scalability of our approach. We argue, however, that this deliberate choice allowed us to gain a deeper understanding of the breadth of the problems by directly engaging with the authors' ideas. Our focus was on comprehending the range of issues, rather than the granular details explored in each paper. The manual review also facilitated our discussion of potential solutions. By directly encountering the developers' words, we were better equipped to formulate our thoughts on addressing the challenges they faced.

\section{Conclusion and Future Work}
\label{sec:Conclusion and Future Work}
This paper delivers a unique two-fold contribution to the Software Engineering and Cyber Human Factors communities. It dives deep into SAST usability challenges, offering valuable insights directly relevant to both SAST tool creators and development teams currently using or considering these tools.

Tool creators can take our results to better inform their development of new and improved SAST tools. Specifically, our findings on developer needs and challenges related to SAST usability can guide the design of more user-friendly interfaces and output display, improved false positive reporting mechanisms, simplified setup, and automatic fixing features. 

Development teams can also leverage our findings to gain insights into potential challenges during SAST implementation. This knowledge can inform their SAST planning process when it comes to false positives, improving tool output and accounting for workflow disruption, ultimately empowering them to use these tools more effectively.

While this study focuses solely on the breadth of how these problems are mentioned, examining the depth at which specific papers discuss them could be valuable future work. This could involve investigating the depth of discussion for each problem and potentially inferring the severity or overall impact it poses for developers. A study of this kind could benefit from some form of NLP to assist in identifying key details in text that signal severity. Our review may also be expanded upon by extending the search back another 5 or 10 years. This could potentially help to gain a broader understanding of the uncovered issues and also identify how their prevalence has changed over time. 

\section{Acknowledgements}
This research is supported by TechLink (TechLink PIA FA8650-23-3-9553). The ChatGPT Large Language Model was used in this paper for spell-checking and grammatical enhancements. A full list of our 89 papers this review was sourced from can be found by following this \href{https://tinyurl.com/bddmj9h8}{\textbf{link.}}


\chapter{Automating Static Code Analysis Through CI/CD Pipeline Integration}\label{CH:manuscript2}

\section{Contribution of Authors and Co-Authors}

Manuscript in following chapter

Author: Zachary Wadhams
\begin{singlespace}
Contributions: Developed study concept and design, data collection and analysis,\\\indent interpretation of results, and wrote the manuscript.
\end{singlespace}

Co-Author: Ann Marie Reinhold
\begin{singlespace}
Contributions: Obtained funding, provided feedback and editing.
\end{singlespace}

Co-Author: Clemente Izurieta
\begin{singlespace}
Contributions: Obtained funding, provided feedback and editing.
\end{singlespace}


\newpage

\section{Manuscript Information}

Zachary Wadhams, Ann Marie Reinhold, Clemente Izurieta

\begin{singlespace}
2nd International Workshop on Mining Software Repositories for Privacy and Security, MSR4P\verb|&|S, (SANER 2024)
\end{singlespace}

\begin{singlespace}
  Status of Manuscript: \\\indent 
  \underline{\phantom{~~X~~}} Prepared for submission to a peer-reviewed journal\\\indent
  \underline{\phantom{~~X~~}} Officially submitted to a peer-reviewed journal\\\indent
  \underline{\phantom{~~X~~}} Accepted by a peer-reviewed journal\\\indent
  \underline{~~X~~} Published in a peer-reviewed journal
\end{singlespace}

\begin{singlespace}
  Publisher: IEEE \\\indent
  In-press
\end{singlespace}

\newpage

\section{Abstract}
In the contemporary landscape of software development, securing sensitive data is paramount to safeguarding organizational reputation, preventing financial losses, and protecting individuals from identity theft. This paper addresses the pervasive challenge of identifying and rectifying security vulnerabilities early in the development process, emphasizing the role of Static Application Security Testing (SAST) tools. While SAST tools play a crucial role in detecting vulnerabilities, widespread adoption has been hindered by usability issues, including high false positive rates and a lack of native pipeline support. This paper proposes a novel, generalized, and automated process for aggregating SAST tool outputs and integrating them into developers' familiar issue-tracking software. The process streamlines the identification and communication of security vulnerabilities during the development lifecycle, facilitating more efficient remediation efforts. We demonstrate the successful implementation of the proposed process with the SonarQube SAST tool in a GitLab-based development environment. Developers were positive about the structured implementation, real-time feedback, and proactive vulnerability management. However, despite some challenges such as a potential learning curve and trade-offs between secure coding and workflow disruption, the overall positive impact on security awareness and responsiveness suggests that the proposed process holds promise in enhancing the security posture of software development practices

\newpage
\section{Introduction}
Now, more than ever, software applications must make a concerted effort to effectively secure the data they store. A single breach of security can wreak havoc on the reputation of the organization, trigger massive financial loss, and even disrupt individuals' lives through identity theft or other means. Many instances of security breaches can be traced back to security vulnerabilities \cite{Thomas2018}. A security vulnerability is a weakness or flaw in a software application's source code or design that can be exploited by malicious actors to compromise the security of the system or the data it processes. It is vital that vulnerabilities are identified and corrected as early as possible in the development process. A key tool that development teams can use to identify vulnerabilities is Static Application Security Testing (SAST), or Static Code Analysis. 

These tools aim to help ensure the security, reliability, and compliance of software applications. SAST tools work by analyzing the source code during the development phase, enabling early detection of security vulnerabilities. By identifying security flaws at an early stage, developers can address them promptly and minimize the risk of such vulnerabilities making their way into the final product. 

SAST tools not only focus on security vulnerabilities but also help improve code quality by identifying design flaws \cite{Jovanovic2006} \cite{Livshits2005}. By detecting and addressing these issues early in the development cycle, teams can enhance the overall quality and maintainability of the codebase. Using these tools, development teams can gain insights into common security pitfalls and improve their understanding of secure coding principles. SAST tools simultaneously serve as educational resources, helping to foster a security-conscious culture among developers who utilize these tools and empowering them to produce code that is more secure and reliable \cite{Nachtigall2022}. 

A small number of such tools are specifically designed to be implemented within Continuous Integration/Continuous Delivery(CI/CD) pipelines on major Git-based repository hosting services. CI/CD pipeline integration allows these tools to run on each new code addition, creating a record (typically stored within the tool GUI or as a .json file) of which changes introduced vulnerabilities or design flaws. The near-immediate feedback given allows for faster remediation of critical vulnerabilities while the historical record can be instrumental in ensuring compliance with a variety of security standards and best practices \cite{Haug2022} \cite{Nguyen2021}. However, the vast majority of tools have little to no support for automation or pipeline integration.

While manual code reviews and standalone static analyses are essential for identifying some design flaws and guiding a project's direction, the complexity, and size of modern software applications make it unfeasible to manually review an entire codebase \cite{Ogheneovo2014}. This necessity calls for automation, which is provided by pipeline integration.

Despite these apparent benefits, many development teams have not deployed SAST tools. Reasons for this vary but many agree that there are still widespread usability issues that hinder their adoption and consistent use. Most of these reasons are commonly identified as high rates of false positives, unhelpful warning messages, lack of fix suggestions, and insufficient native pipeline support \cite{Ayewah2010}. Developers often delegate security-related concerns to colleagues within their organization or, in some cases, they may lack access to these reports due to security configurations \cite{Lopez2023} \cite{Xie2011}. As those possessing the most profound understanding of the code, developers often overlook security issues, which can lead to the ``out of sight, out of mind" problem \cite{Lopez2023} \cite{Nachtigall2022}.

Instead of merely investigating this issue, we have chosen to design a solution that involves presenting SAST reports to developers in a familiar and consistent manner using automated methods. While many have touched upon this issue, we have identified no papers that discuss a generalized process applicable to all static analysis tools. By offering a standardized process that is easy to integrate and demonstrating the tangible benefits of doing so, we aspire to promote the widespread adoption and integration of static analysis tools within the software development community. This, in turn, enhances software privacy and security across a broad spectrum of domains and applications. 

Our first key contribution is a \textbf{generalized and automated} process for aggregating SAST tool outputs and integrating them into developers' familiar issue-tracking software. This process streamlines the identification and communication of security vulnerabilities during the development lifecycle, facilitating more efficient remediation efforts.

In addition, we provide a detailed examination of a practical implementation of the proposed process. By illustrating its application in a real-world scenario, we offer insights into the feasibility and effectiveness of the process. The practical implementation serves as a valuable case study, shedding light on the challenges encountered and the practical considerations that arise when implementing such a process within software development teams.

Our work contributes to the security vulnerability management field by thoroughly exploring the potential benefits and drawbacks of the explored process. By addressing advantages such as improved collaboration between security and development teams, which in many cases are separated by technology and/or organizational constraints, our aids in understanding the implications and trade-offs associated with adopting our process.

Importantly, these contributions hold direct relevance to the Mining Software Repositories (MSR) community. The proposed process aligns with the objectives of MSR by leveraging data from software repositories, issue-tracking systems, and security testing tools to better inform software developers of potential security vulnerabilities and foster proactive measures for secure software development practices, ergo the ability to mine software repositories to make information available to potentially separate engineering groups (i.e., security and development) in a CI/CD environment is critical. The insights gained from this research can inform and enhance the broader understanding of how security practices intersect with the larger software development ecosystem, thereby contributing to the ongoing advancements in the MSR field.

This paper is structured as follows: The Motivation section explains the motivation for our work in the field of static analysis. The Related Work section discusses related work and its impact on our research. The Process section describes our process and its core components in detail. The Use Case section presents a practical example of an implementation of our process. The Discussion section provides a discussion on the greater consequences of our work. The Threats to Validity section addresses potential threats to the validity of our work and our attempts to mitigate them. Section The Conclusion and Future Work section concludes the paper and outlines our plans for future work on refining and extending the proposed SAST tool integration process.

\section{Motivation}
\label{sec:Motivation}
It is the moral and now legal obligation of organizations to ensure that any and all software they release is as safe and secure as possible and respects the privacy of its users. The EU's General Data Protection Regulation (GDPR) is just one example of legislation which requires that organizations ensure privacy and security are built into their applications by design \cite{GDPR2016}.

SAST tools excel in this area where other methods, such as dynamic analysis, fall behind. Dynamic analysis is performed later in the development process, typically when an application is executed. Static analysis strictly looks at source code and can be implemented as soon as the first line of code is written. This reinforces SAST's use as a proactive step that can be used early in the development process when an application's design is still flexible, bringing it in line with the core tenet of privacy and security by design. Although static analysis procures many benefits, it is often utilized inconsistently.

Issues with organizational security configurations can make the consistent use of SAST difficult for some. Occasionally, organizations restrict the execution of tools, requiring them to be behind a firewall and rendering them inaccessible to certain developers. This becomes a problem when those with access are senior developers who do not have time to sift through the reports generated by tools, leading to slow turnaround times on fixes or even vulnerabilities being missed completely. Thus, a disconnect exists between the development teams and security experts or senior developers who are directly responsible for ensuring code security.

This disconnect demands a better way to bring more average developers into the sphere of security and subsequently motivated our team to design a process that does just that. Leveraging the familiarity of bug and issue-tracking features within popular repository hosting services, the goal is to increase the visibility of SAST tool outputs, ultimately providing a consistent and accessible space for identifying, managing, and resolving design flaws and vulnerabilities within software repositories. 

\section{Related Work}
\label{sec:Related Work}
Interest in SAST has been steadily growing over the past two decades, while interest in security and privacy by design has burgeoned since the publication of the GDPR in 2016 \cite{GDPR2016}. The combination of SAST tools and the concept of privacy and security by design, and how one can drive the usage of the other, inspired our work on the topic.

Johnson et al. \cite{Johnson2013} carried out interviews to investigate why static analysis tools were not being used by developers. During their research, they uncovered many underlying issues with these tools, such as poor output presentation and slow feedback.  In other work performed by Izurieta et al. \cite{Izurieta2013}, the uncertainty of scoring and error propagation from SAST tools is also addressed.

Haug et al. \cite{Haug2022} showed that when developers are provided with immediate or near-immediate feedback on code, they are more likely to consider it.

Xie et al. \cite{Xie2011} found that most software security vulnerabilities are caused by errors introduced by developers. Their results led them to discover a striking divide between developers' security knowledge and their practices. They concluded that static analysis tools do play an important role in assisting developers in producing secure software and overcoming their apprehension towards security.

Ayewah et al. \cite{Ayewah2010} observed that when developers are present with security-related messages, they generally make the correct decisions to address them even with little to no formal security experience.

Nachtigall et al. \cite{Nachtigall2022} conducted a comprehensive study focused on specific criteria for static analysis tools. They analyzed 36 criteria across 46 different tools from a user's perspective. The study revealed significant shortcomings in many SAST tools, particularly concerning their integration into developers' workflows. It was identified that the locations where tool outputs are stored is unfamiliar to many developers and accessing them often requires them to change their habits. In some cases, these tools have their data stored behind firewalls that only senior developers or cybersecurity officers can access. Whether intentional or not, this leaves other developers in the dark about the potential vulnerabilities they may introduce, thereby compromising the security culture of the organization. Many tools also provide these outputs in unformatted text that doesn't grab a developer's attention. As a result, they found that if a tool disrupts a developer's workflow or lacks sufficient visual guidance, many will quickly abandon it. This underscores a critical and recurring usability challenge. 

Our process not only addresses these previously identified issues but also establishes a way for delivering static analysis tool outputs to developers in a non-disruptive and visually appealing format, ultimately ensuring that a broader range of developers can easily access these reports, thereby enhancing, rather than impeding, their productivity and workflow.

\section{Process}
\label{sec:Process}
We implement our process in three steps. First, we identify the SAST tools preferred by the organization and assess the types of data they provide. Two, we explore the development environment of the target organization, taking into account aspects such as issue-tracking software, repositories, pipeline configuration, and network security configurations. Three, we implement a controller script that ties the SAST tools and issue tracking together. The sub-sections below provide a detailed explanation of each of these steps.

\begin{figure}[t]
\centerline{\includegraphics[scale=0.45]{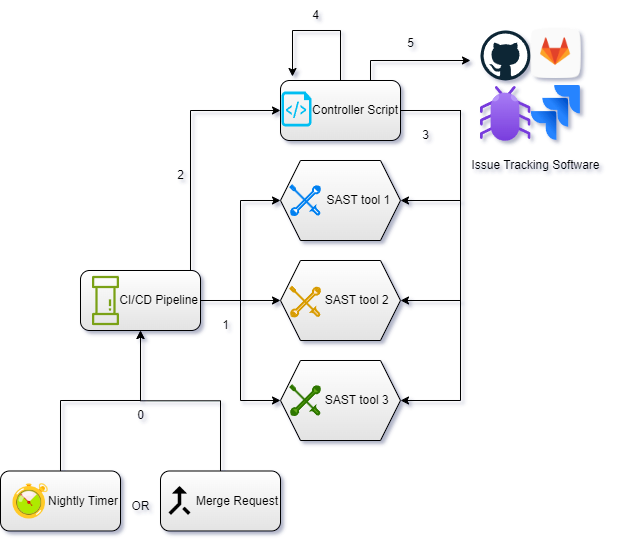}}
\caption{ Process flow diagram. \\ \\ The boxes represent important concepts or technologies while the arrows depict the flow of the process. In step \textbf{0}, some external factor, such as a nightly timer or a developer-initiated merge request, triggers the build pipeline. The pipeline then initiates the analysis of each SAST tool in step \textbf{1}. Once all static analyses are completed, the build pipeline starts the controller script in step \textbf{2}. The controller script reaches out to each SAST tool and gathers the relevant issue data in step \textbf{3}. The issue data is then formatted by the controller script and assembled into payloads in step \textbf{4}. In step \textbf{5}, each payload is sent to the issue tracking software, and individual issues are created.}
\alt{Process flow diagram. The boxes represent important concepts or technologies while the arrows depict the flow of the process. In step 0, some external factor, such as a nightly timer or a developer-initiated merge request, triggers the build pipeline. The pipeline then initiates the analysis of each SAST tool in step 1. Once all static analyses are completed, the build pipeline starts the controller script in step 2. The controller script reaches out to each SAST tool and gathers the relevant issue data in step 3. The issue data is then formatted by the controller script and assembled into payloads in step 4. In step 5, each payload is sent to the issue tracking software, and individual issues are created.}
\label{figure1}
\end{figure}

\subsection{Tool Identification and Data Assessment}
Diverse organizations have unique analytical requirements, prompting the need for discussions on preferred static analysis tools. These tools can be categorized into two groups: those with a Graphical User Interface (GUI) and those with a Command Line Interface (CLI). Our process focuses on GUI tools. They typically have the most customizability and often offer an Application Programming Interface (API). The API simplifies data retrieval by returning it seamlessly in commonly formatted structures such as XML, JSON, or HTML. Our API-driven process enhances the efficiency of data extraction, making it easier to integrate the data into our controller script. 

In our context, we treat GUI tools without APIs the equivalently to CLI tools. CLI tools are executed exclusively from the command line and do not provide the additional functionality and ease of use offered by an API. After completion, these tools generate a file formatted as either XML or JSON containing the report data. While CLI tools can provide valuable information, our process does not focus on them.

After identifying which tools the organization desires, we conducted an investigation of the organization's familiar development environment.

\subsection{Development Environment Exploration}
The development environment of the target organization is taken into account when custom tailoring the implementation of our process. We considered what software most organizations use to manage their codebase and found that the vast majority use one of three Git-based systems; GitHub, GitLab, and BitBucket. Many development teams either utilize the built-in repository issue tracking or rely on connected software adjacent to the repository for issue management \cite{Baysal2014}. Each of these services provides a robust API to assist with automating aspects such as bug and issue tracking. These APIs allow for calls to be made that create, edit, or resolve issues. An ``issue" or sometimes ``ticket" is a digital record used to track tasks, bugs, and feature requests related to a software project. They help teams collaborate by providing a place to discuss, assign, and monitor the progress of these work items. APIs are ubiquitous in issue tracking, enabling this process to be applied to any service through the controller script.

The build pipeline of the repository is a predefined and automated sequence of tasks and actions that transform source code into a deployable application or software artifact. This pipeline is where the code is compiled and also the stage at which SAST tools are executed. The CI/CD pipelines of the aforementioned repository hosting services are standardized through the use of a common file type used to choreograph the execution steps of the pipeline. The controller script is placed after the SAST tool execution in the pipeline order to ensure that each analysis is completed before any data is retrieved. 

As previously mentioned in the Motivation Section, some organizations may have security configurations that restrict the execution of tools or the CI/CD pipeline behind a firewall, rendering them inaccessible to certain developers. Our process accommodates such security controls while allowing for developer access, as long as the machines running the tools and the pipeline can be configured to communicate with each other and bypass firewall rules. Some organizations may cite vulnerability data as a potential security concern that should not be shared widely. However, the developers who would be fixing security vulnerabilities must already have access to the project's source code. This source code access, combined with the fact that many SAST tools are open source and free to use, means that anyone with source code access could run an analysis of their own and obtain these reports if they wished to, making the reasoning behind this security concern unsound.

After collecting the necessary tools and information about the development environment, the controller script can take shape.

\subsection{Controller Script}
\label{sec:Controller Script}
The controller script serves as the keystone that seamlessly connects the previously disjoint processes of SAST report generation and issue management. Any scripting language is acceptable as long as it is supported by the hosting pipeline. Its primary role is to orchestrate the flow of data between the SAST tools and the issue tracking system, ensuring a smooth and automated transition. This bridge is established through a series of well-defined steps. 

The controller script begins by collecting the SAST report data generated by whatever tools the target organization chooses to use (Figure \ref{figure1}, step 3).  Through the tool's API, raw SAST data is obtained, and the controller script transforms it into a standardized format that aligns with the requirements of the issue tracking system (Figure \ref{figure1}, step 4). At a minimum, the format includes a title that identifies the issue type and severity, a one-sentence description of the problem, an identifier indicating the specific line of code and file where the issue resides, and either a problem description if provided by the tool or a unique reference number such as Common Vulnerabilities and Exposures (CVE) or Common Weakness Enumeration (CWE). This transformation ensures that data from diverse tools is uniform and can be consistently integrated into the tracking system. Before sending the data as an issue to the tracking software, it must be converted to markdown, ensuring compatibility and consistency with the tracking system's formatting and requirements.

Leveraging the issue tracking system's API  the controller script automates the creation (Figure \ref{figure1}, step 5) of issues or tickets to report vulnerabilities, weaknesses, or other code-related concerns identified by the SAST tools. It establishes near real-time synchronization through the pipeline, updating the issue tracking system with the latest analysis results whenever the pipeline runs due to developers committing code (Figure \ref{figure1}, step 0).

To prevent the creation of duplicate issues within the issue tracking software, issues within the SAST tool must be marked. One straightforward process is to update the status of all issues that the script has identified and 'moved' within the tool. This ensures that upon each new analysis, issues that have already been seen by the script are ignored and only new issues are considered (Figure \ref{figure1}, step 3). 

In many cases, it may be valuable for the script to include a quality gate. Quality gates play a crucial role in maintaining the integrity of the software development process. They function by continuously monitoring the code changes as they pass through the CI/CD pipeline. If the gate detects the introduction of major vulnerabilities or other high-risk issues, it will immediately halt the pipeline's progress after generating the issues, alerting developers of a potential problem. This preventive measure ensures that no changes with severe flaws are allowed to proceed further into the development or deployment stages.

In practice, a quality gate serves as an additional layer of defense, reinforcing the security and quality of the software. It ensures that any code changes are thoroughly examined for major vulnerabilities before they can proceed, thereby contributing to a more robust and reliable software development process.

\begin{figure}[t]
\centerline{\includegraphics[scale=0.55]{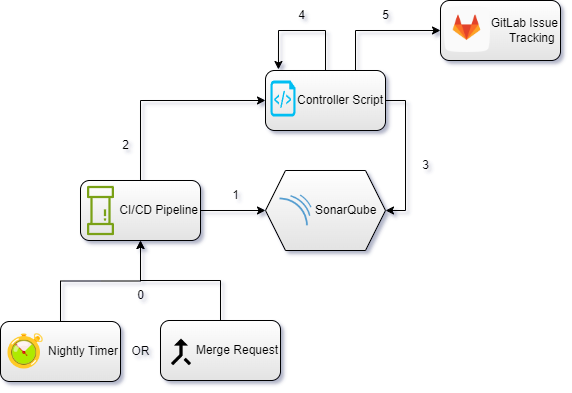}}
\caption{Implementation of Approach to an Example Organization (Organization X). \\ \\ The boxes represent important concepts or technologies while the arrows depict the flow of the process. The implementation begins with either a nightly timer or merge request in step \textbf{0}. The pipeline then initiates the analysis SonarQube in step \textbf{1}. Once SonarQube's analysis is complete, the build pipeline starts the controller script in step \textbf{2}. The controller script reaches out to SonarQube and gathers the relevant issue data in step \textbf{3}. The issue data is then formatted by the controller script and assembled into payloads in step \textbf{4}. In step \textbf{5}, each payload is sent to GitLab's issue tracking software, and individual issues are created therein.}
\alt{Implementation of Approach to an Example Organization (Organization X). The boxes represent important concepts or technologies while the arrows depict the flow of the process. The implementation begins with either a nightly timer or merge request in step 0. The pipeline then initiates the analysis SonarQube in step 1. Once SonarQube's analysis is complete, the build pipeline starts the controller script in step 2. The controller script reaches out to SonarQube and gathers the relevant issue data in step 3. The issue data is then formatted by the controller script and assembled into payloads in step 4. In step 5, each payload is sent to GitLab's issue tracking software, and individual issues are created therein.}
\label{figure2}
\end{figure}

The versatility of the controller script is a fundamental strength of our process, as it can be tailored to collect diverse types of issues such as vulnerabilities, bugs, or design issues while filtering them by potential severity or impact. The script can be configured to recognize and process different issue categories by adjusting the data collection and transformation steps, thereby accommodating the specific needs and priorities of the development team. By integrating these functions, the controller script effectively bridges the gap between the security-focused static analysis and the broader software development process, fostering an efficient and proactive process to addressing code vulnerabilities and design concerns.

\section{Use Case}
\label{sec:Use Case}
To test the practical application of our process we worked with an organization whose goal was to implement a static analysis tool to enhance the security of their in-development application. To protect this organization's identity, we refer to them as Organization X. 

Figure \ref{figure2} depicts Organization X's implementation of our process, following the same steps outlined in Figure \ref{figure1}.

Being the first static analysis to be used on their repository, a tool that is simple to configure but still offers customizability and scales well with a rapidly growing codebase was deemed necessary. Following our outlined process, Organization X selected SonarQube as their preferred tool after evaluating its features and usability. 

The development environment of Organization X consisted of a GitLab repository where code is hosted, the pipeline is managed, and issues are tracked (Figure \ref{figure2}, steps 0, 5). A key facet to note is that the repository build pipeline as well as all execution of external tools were required to be placed behind a firewall to comply with the organization's security requirements. This was a hard requirement that, as a consequence, restricted who could work with SonarQube to two senior developers out of more than twenty total developers. As SonarQube stores outputs within its GUI, this issue exemplifies the problem discussed in section II (i.e., Motivation) where some developers are segregated due to accessibility of sensitive data. In the absence of a structured process, only two developers would be able to configure, maintain, and then be responsible for the potentially very large number of vulnerabilities and design flaws that could be uncovered. 

\begin{figure}[t]
\centerline{\includegraphics[scale=0.55]{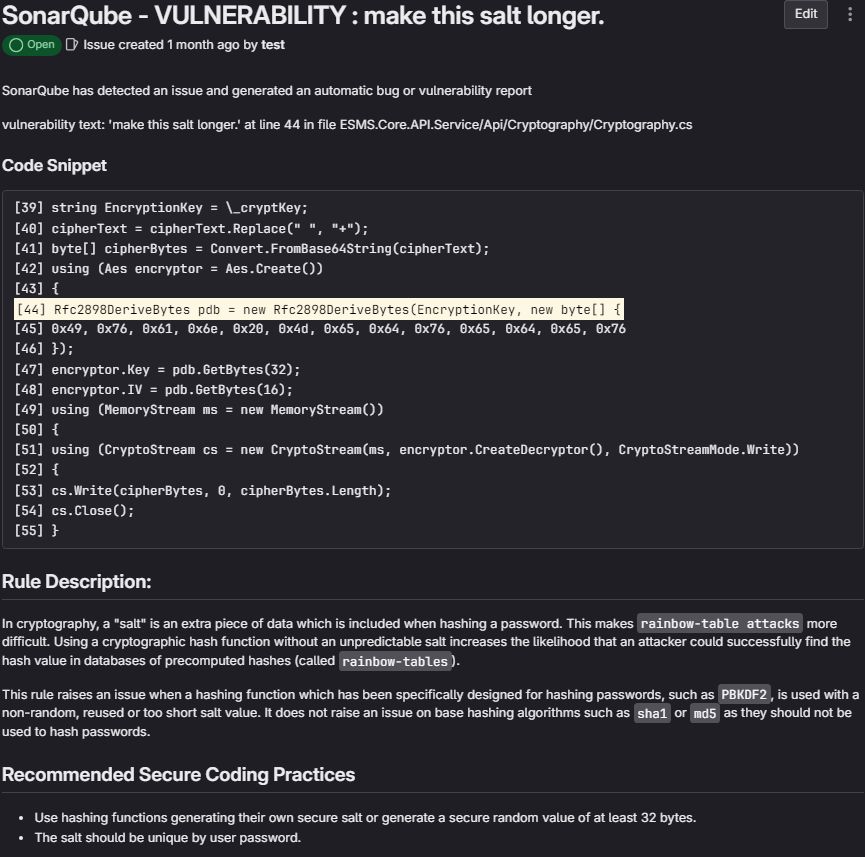}}
\caption{Example of a Generated Issue}
\alt{Example of a Generated Issue}
\label{figure3}
\end{figure}

Organization X used our process, implementing the controller as a Python script (Figure \ref{figure2}, step 2). This script utilized the Requests package to make HTTPS requests to the SonarQube and GitLab instances. SonarQube has the capability to uncover three different types of issues in projects: bugs, vulnerabilities, and code smells. Organization X was interested only in vulnerabilities and bugs, so these requests were targeted to endpoints within SonarQube's Web API to retrieve data related to those types of issues (Figure \ref{figure2}, step 3). SonarQube also attaches a severity to each issue it finds. Organization X decided to consider only bugs and vulnerabilities with a severity of critical, major, or high in an effort to triage issues that have a larger potential to cause problems. 

After retrieving the data, the script extracted the essential information from the JSON objects. Subsequently, a payload was assembled and sent to GitLab's issue-tracking software (Figure \ref{figure2}, step 4). Organization X decided on the data that would be most relevant and useful to their developers, choosing to include selected information in generated issues. The one-sentence description of the created GitLab issue comprised the SonarQube issue type, title, and severity. The body of the issue is generated with a disclaimer, stating that this issue was automatically created using data from SonarQube. Organization X considered this necessary to help their developers differentiate between automatically generated and manually created issues. The rest of the issue body contained a code snippet that showed the problem line of code along with the surrounding 10 lines for context, an explanation of why the issue should be addressed, and a link to the corresponding CWE or CVE for further reading. Each payload was then sent to GitLab's API as a create issue request and would appear alongside developer-created issues within GitLab's issue tracking software for developers to pick up and address (Figure \ref{figure2}, step 5).

As this process was tested, starting with Organization X's nightly pipeline runs, they found it to be helpful as it didn't disrupt their established pipeline and did not add a significant amount of runtime. Developers also reacted positively to the formatting and location of the issues, commenting that the problems in the code were easily identifiable, and the additional information provided aided them in engineering fixes. Figure \ref{figure3} depicts an example issue that Organization X generated in GitLab using data from SonarQube.

After testing for a few months and being satisfied with the results, Organization X decided to develop a secondary controller script designed to run on their merge request pipeline with every developer code change. The aim of this was to prevent developers from introducing vulnerabilities and bugs through the addition of a quality gate. When a developer attempts to merge their code with the main code branch, the quality gate checks whether the code changes would introduce bugs or vulnerabilities. If they do, the changes are rejected, and the developer is notified to fix the issues before merging their code. If the changes do not introduce bugs, the merge proceeds as normal. Organization X chose to reject changes if they introduce more than one blocker vulnerability or bug or more than two critical vulnerabilities or bugs.

\section{Discussion}
\label{sec:Discussion}
The developer feedback provided us with insights into how they viewed different aspects of the implementation of SAST tools into their workflow using our process. For example, when the quality gate was implemented, a few violations were noted on the first day of use. The gate functioned as expected, rejecting the changes. Feedback on the gate from developers was conflicting; while all understood its necessity, opinions varied on its rigidity. Some opposed the aggressive process, suggesting issues be noted for later resolution to allow focused coding. Others appreciated it, foreseeing time saved in the long run. This pinpointed the quality gate as a potential challenge in terms of the developer's workflow and productivity.

Developers appreciated the structured implementation of the process, expressing satisfaction with the automation of vulnerability and bug identification while positive sentiments were shared about the seamless integration with GitLab's API and the existing pipelines. Perhaps the most commonly identified upside to the automated generation of issues was the real-time feedback, which either helped developers immediately fix the issues or begin planning to address them at a future date. Developers acknowledged the significant enhancement in code security through systematic vulnerability identification and positively recognized the proactive process of the quality gate in effectively managing vulnerabilities. Over a short period, numerous previously unknown issues were uncovered. While some of these turned out to be false positives, others were genuine and had been lurking within the codebase for months. Organization X's developers stated that without the implementation of static analysis, these issues might have gone undetected. As developers are exposed to more vulnerabilities, there is potential for a relief of tension between development and security teams through increased collaboration. 

However, our process was not without complications. A potential problem developers identified was that due to the formatting, identifying false positives required more effort. Developers had to delve deeper into an issue before realizing that it was a non-issue. This could be argued as a trade off for having issues formatted in a detailed way. The introduction of the new process faced initial resistance as it presented a learning curve to developers who were already comfortable with existing practices.

The developer feedback we received illuminates important questions to the MSR community. What is the value of the trade-off between having secure code and the potential disruption to a developer's workflow? How can a balance between those two important aspects be achieved? It is evident that while the benefits of the automated generation of issues are substantial, addressing the initial resistance and facilitating a seamless adaptation process are key aspects to consider in the ongoing refinement of the process. 

\section{Threats to Validity}
\label{sec:Threats to Validity:manuscript2}
Our study contributes valuable insights to the fields of SAST, vulnerability management, and MSR. Here we address potential threats to the validity in our study. By identifying and acknowledging these limitations, we aim to provide a transparent assessment of the scope and generalizability of our work. This section outlines key threats to the validity of our study, offering a comprehensive view of the factors that may impact the reliability and applicability of our results.

First, while we attempted to make our process as generalized as possible, Organization X may have unique characteristics that limit the generalizability of findings to other organizations. To address this, we thoroughly explored their development environment and compared it to other known development environments. While hardly exhaustive, we identified no known out-of-the-ordinary development practices. Second, the definition and identification of vulnerabilities and bugs may vary among different SAST and their versions \cite{Reinhold2023}. Since SonarQube was the sole tool utilized by Organization X, we lack a practical implementation of other tools to compare our results. Addressing this limitation is part of our future work, as detailed in the following section, where we plan to test with additional tools to confirm the generalizability of our process.

\section{Conclusion and Future Work}
\label{sec:Conclusion and Future Work:manuscript2}
In this paper, we have addressed the critical issue of effectively integrating SAST tools into the software development lifecycle. This work's noteworthiness to the MSR community cannot be understated due to its direct alignment with the overarching goals of deriving actionable information relating to privacy and security. The increasing importance of securing data and the potential consequences of security breaches highlight the necessity for proactive measures in identifying and addressing security vulnerabilities. Our process focuses on automating the aggregation and integration of SAST tool outputs into developers' familiar issue-tracking software, thereby enhancing the visibility and accessibility of security-related issues. Our results from the use case suggest that developers are mostly satisfied with the way these issues are presented and that the addition of them to their workflow has not been overly intrusive.

While our results are encouraging, there is still much work to be done on the our procedure, such as adding additional functionality and conducting use cases with other tools. For instance, many SAST tools are CLI-only and lack an API for data retrieval. In the future, we will address these tools as well by incorporating them into our process, rather than solely focusing on large web-based tools. Some of these CLI tools provide little to no detail on the issues they uncover, except for a reference to their corresponding CWE or CVE. We will investigate whether retrieving data from the CWE/CVE databases to fill out more information in these issues would be valuable. Additionally, we plan to conduct other use cases that follow our process with different SAST tools, potentially incorporating more than one tool at once. 

\section{Acknowledgements}
This research is supported by TechLink (TechLink PIA FA8650-23-3-9553). The ChatGPT Large Language Model was used in this paper for spell-checking and grammatical enhancements. 

© 2024 IEEE. Reprinted, with permission, from Z. Wadhams, A. Reinhold, C. Izurieta., "Automating Static Code Analysis Through CI/CD Pipeline Integration," in 2024 IEEE International Conference on Software Analysis, Evolution and Reengineering 2nd International Workshop on Mining Software Repositories for Privacy and Security, Rovaniemi, Finland: IEEE, March 2024. [In-press].
\chapter{Conclusion}\label{conclusion}

This research has presented two interconnected manuscripts that collectively contribute to a deeper understanding of the challenges developers face when utilizing SAST tools and potential solutions to enhance their usability. 

The first manuscript explores the ever-expanding field of SAST tools and implementations. This in-depth study analyzed 89 research papers (Metric 3) to identify common challenges faced by software developers when using SAST tools. The list of problems (Metric 1), accompanied by the frequency of their mention in the literature (Metric 2), provides valuable insights to readers. As mentioned in the manuscript, this research is dual-purpose, benefiting both software development teams and tool creators. It informs developers who may be considering adopting SAST about what they can expect. Simultaneously, it outlines problems that SAST tool creators can work towards improving with new tools or updates to old ones. 

The manuscript also looks deeply into the human factors that influence SAST utilization. A very wise person once said to me, no matter how effective a seatbelt is, it's useless if no one chooses to wear it. This is precisely the issue with SAST. We know that these tools can have a positive impact on the development process, but if the perceived drawbacks outweigh the perceived benefits, they will be quickly abandoned.

The second manuscript, ultimately, is an attempt to challenge and change this balance of perceived drawbacks and perceived benefits. The process shown in Figure \ref{figure1} was designed with generalizability in mind. No two development environments are the same; they all have their own unique niceties. These niceties are often what developers cling to and have come into existence to solve problems specific to their team. Perhaps your team has been using GitHub Issues to track your bugs and vulnerabilities for years. You are not going to want to change your issue tracking software for the sake of adopting a SAST tool. Trying to change these features and forcing conformity was specifically avoided, as I believed that a non-generalized process would be doomed to fail.

To evaluate how this process performed, only a real world implementation would provide the feedback I sought. The Organization X use case discussed in the manuscript provided this feedback (Metric 4). As noted in the manuscript, the feedback included some criticisms but overall there was a positive shift in the perceived benefits of the implemented SAST tool. Hopefully, this work can motivate others to challenge that balance in a more aggressive way. Some may push developers too far and fail, but others may eventually change the status quo in a larger and more meaningful way. 

Software security is paramount to the integrity and functionality of our modern society. In today's interconnected world, we can no longer afford to neglect software security and quality. It must be a top priority for everyone involved in the development process, from initial design to final testing and deployment to long term maintenance.

This thesis, through the two manuscripts within, provides valuable insights to the field of SAST for software security. It shows that there is an issue with the adoption and continued use of these tools and that developers can be apprehensive to use something that may disrupt their workflow or that hasn't had the chance to earn their trust. Then, the thesis goes further by proposing a process to directly address some of these uncovered issues, specifically workflow integration and tool automation improvements. This thesis contributes to the advancement of SAST by providing a comprehensive analysis of its current adoption landscape. The proposed recommendations offer practical guidance for organizations seeking to optimize their use of SAST tools and mitigate associated risks. By addressing the identified challenges, this research paves the way for a more secure and resilient software ecosystem.


\bibliographystyle{plainnat} 
\references
\AtBeginEnvironment{thebibliography}{\linespread{1}\selectfont}
\bibliography{Cumulative} 

\end{document}